\documentclass[runningheads]{llncs}
\usepackage{amsmath,graphicx}
\usepackage{amssymb}
\usepackage{multirow}
\usepackage{mathrsfs}
\usepackage{amsfonts} 
\usepackage{hyperref}
\usepackage{booktabs} 
\usepackage{multirow}
\usepackage{xcolor}
\begin{document}

%
\title{Reconstruction and Quantification of 3D Iris Surface for Angle-Closure Glaucoma Detection in Anterior Segment OCT}
%
%
\author{Jinkui Hao\inst{1,6}
\and Huazhu Fu\inst{2*}
\and Yanwu Xu\inst{3}
\and Yan Hu\inst{4}
\and Fei Li\inst{5}
\and Xiulan Zhang\inst{5}
\and Jiang Liu\inst{4}
\and Yitian Zhao\inst{1*}
}



\institute{Cixi Institute of Biomedical Engineering, Ningbo Institute of Materials Technology and Engineering, Chinese Academy of Sciences, Ningbo, China, \email{yitian.zhao@nimte.ac.cn}
\and Inception Institute of Artificial Intelligence, \email{hzfu@ieee.org}
\and Baidu Inc
\and Southern University of Science and Technology
\and State Key Laboratory of Ophthalmology, Zhongshan Ophthalmic Center, Sun Yat-sen University, Guangzhou
\and University of Chinese Academy of Sciences, Beijing, China
}

\authorrunning{Jinkui Hao et al.}

\maketitle              
\begin{abstract}

Precise characterization and analysis of iris shape from Anterior Segment OCT (AS-OCT) are of great importance in facilitating diagnosis of angle-closure-related diseases. 
Existing methods focus solely on analyzing structural properties identified from the 2D slice, while accurate characterization of morphological changes of iris shape in 3D AS-OCT may be able to reveal in addition the risk of disease progression.  
In this paper, we propose a novel framework for reconstruction and quantification of 3D iris surface from AS-OCT imagery. We consider it to be the first work to detect angle-closure glaucoma by means of 3D representation. 
An iris segmentation network with wavelet refinement block (WRB) is first proposed to generate the initial shape of the iris from single AS-OCT slice. 
The 3D iris surface is then reconstructed using a guided optimization method with Poisson-disk sampling.  
Finally, a set of surface-based features are extracted, which are used in detecting of angle-closure glaucoma. 
Experimental results demonstrate that our method is highly effective in iris segmentation and surface reconstruction. Moreover, we show that 3D-based representation achieves better performance in angle-closure glaucoma detection than does 2D-based feature.

\end{abstract}
\section{Introduction}

\begin{figure}[!t]
\includegraphics[width=12cm]{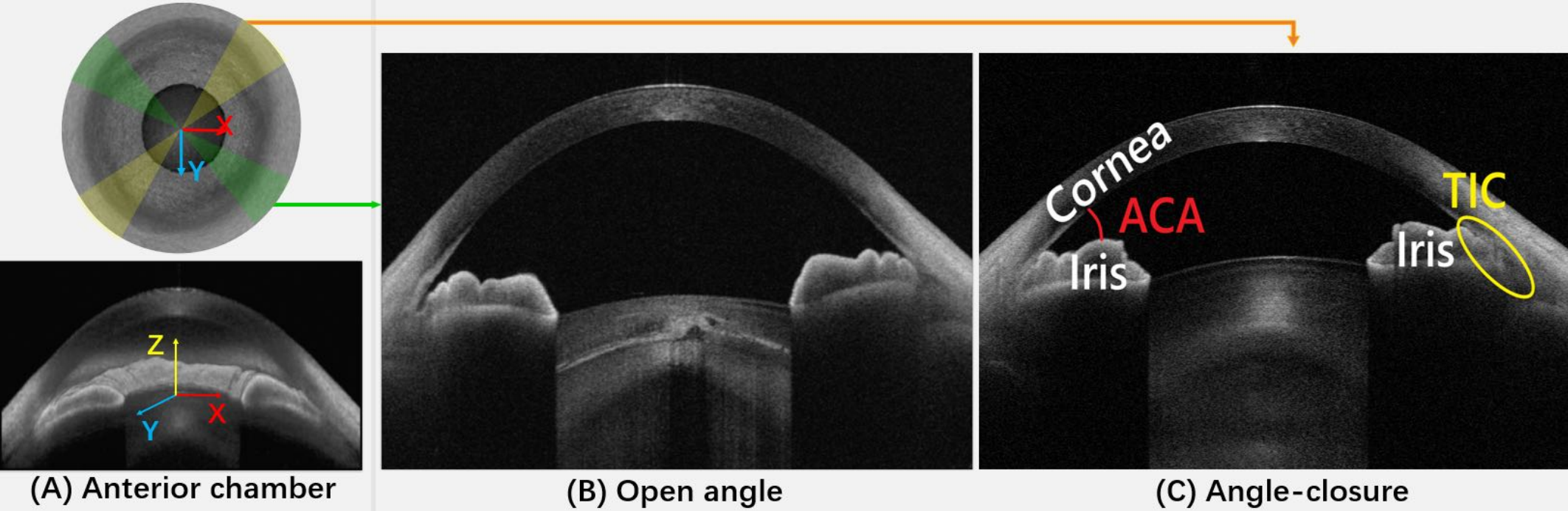}
\centering
\caption{For one AS-OCT volume (A), the open angle (B) and angle-closure (C) cases may appear in different sectors. A 3D volume is more suitable than a 2D image for supporting global analysis.  } \label{fig1}
\vspace {-12pt}
\end{figure}

Anterior Segment OCT (AS-OCT) imaging is a non-contact and non-invasive method for cross-sectional viewing of anterior segment structure, as shown in  Fig.~\ref{fig1} (A). Anatomical structures, such as iris shape and anterior chamber angle (ACA), observed in AS-OCT play key roles in facilitating examination and diagnosis of angle-closure glaucoma~\cite{Chansangpetch2018,Ang2018,Xu2019_AJO}. Fig.~\ref{fig1} (B, C) show two AS-OCT images revealing  open angle and angle-closure glaucoma, respectively.
However, manual identification of angle-closure glaucoma is time consuming and prone to human error.
To this end, automated extraction of morphological features, e.g., ACA, iris and other anterior segment structures, would  benefit both  clinical diagnosis and any automated screening system~\cite{fu2017segmentation,Fu2018_MICCAI,Fu2019_AJO,Fu2019TCb}.

Epidemiological studies~\cite{wang2010quantitative,huang2015comparison} have established that quantitative iris parameters are independently related to narrow ACA, and an anteriorly-bowed iris may be related to the degree of angle-closure progression.  Huang et al.~\cite{huang2015comparison} also
suggest that morphological changes in the iris surface are an important sign, revealing and enabling the understanding of the pathogenesis of angle-closure glaucoma. 
As a result, automated extraction of the iris from the AS-OCT has become an active research area of significance for future diagnosis and prognosis.
Ni et al.~\cite{ni2014anterior} assessed angle-closure glaucoma by computing mean iris curvature and the trapezoidal area of the iridocorneal angle, etc. Fu et al.~\cite{fu2017segmentation} proposed a data-driven method of segmenting the cornea and iris, as well as measuring the clinical parameters essential to screen for glaucoma. Shang et al.~\cite{shang2019tensor} presented a curvilinear structure filter based on the local phase tensor to extract the iris region, so as to further assist the diagnosis of angle-closure glaucoma.
However, all of the aforementioned methods rely on 2D slices of AS-OCT, which are less useful in distinguishing the stages of angle-closure glaucoma. This may stem from the fact that AS-OCT provides only a single cross-sectional slice view across the anterior segment and, in consequence, all other slices are irrelevant to the task of determining angle status~\cite{cho2017evaluation}. In contrast, a comprehensive study of the global information provided by a 3D representation of the iris may improve  measurement accuracy and robustness more significantly than conventional approaches that make use of only an individual 2D slice. Moreover, the occludable iridocorneal, or fully closed ACA leads to the presence of trabecular iris contact (TIC) and exacerbate the iris reconstruction problem.

To this end, in this paper we propose an automated reconstruction and quantification framework for 3D iris surface. 
Inspired by the discrete wavelet transform, we introduce a novel wavelet refinement block (WRB) into a U-shaped architecture~\cite{ronneberger2015u} with a view to reducing the redundancy while maintaining local details to the decoder, for extracting an initial iris segmentation. The detected iris boundaries are then utilized to reconstruct a 3D iris surface based on  adapting an Poisson-disk sampling. Finally, we extract features from this 3D iris surface (e.g., principal curvatures, Gaussian curvature, mean curvature and shape index) to further assist the examination and diagnosis of angle-closure glaucoma. The experiment demonstrate that our proposed method has high effectiveness on iris segmentation and reconstruction. 

\begin{figure}[t]
\includegraphics[width=\textwidth]{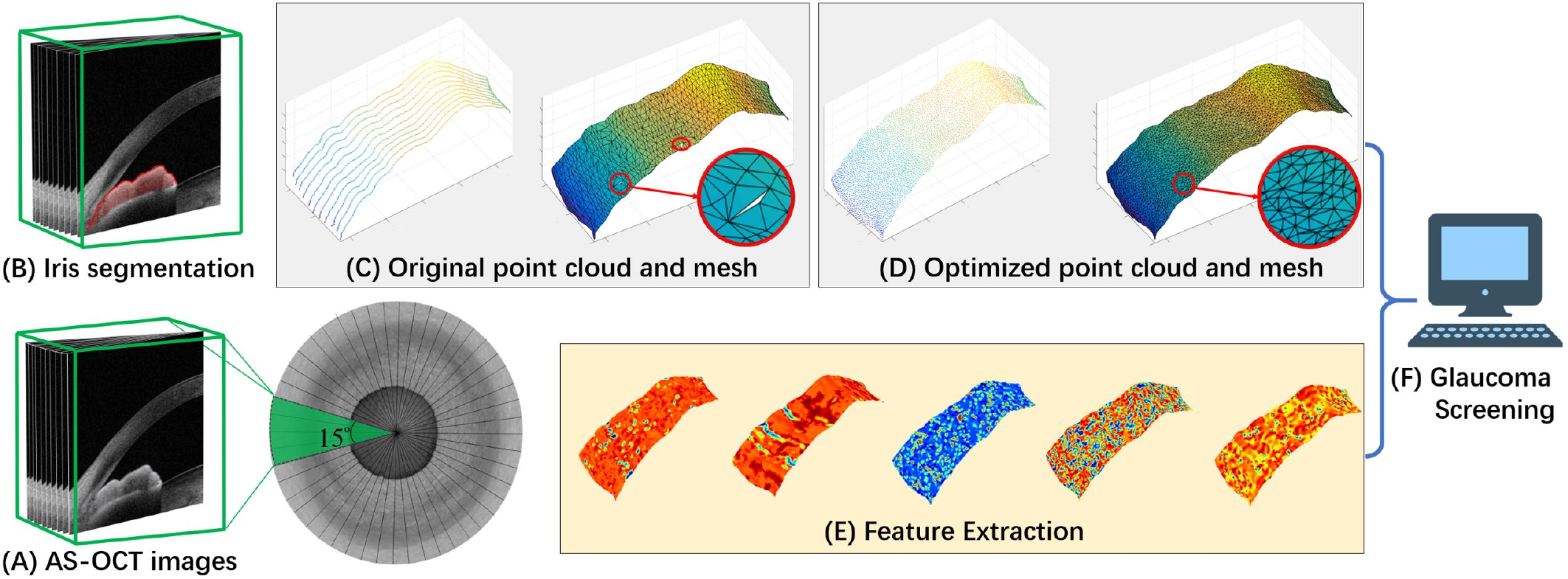}
\caption{Flowchart of the proposed approach. Given an AS-OCT sector in a $15^{\circ}$ radiant area (A), the iris boundaries (B) are firstly identified by our segmentation network. We then convert the segmentation into a 3D point cloud and mesh (C). A constrained Poisson-disk sampling method is used to optimize the point cloud and mesh (D), so as to obtain a more accurate iris surface.  Finally, different surface measurements (E) are computed for diagnosis of angle-closure related diseases (F). } \label{fig2}
\vspace{-12pt}
\end{figure}

\section{Proposed Method}

In this section, we introduce the proposed 3D iris reconstruction and quantification framework for angle-closure glaucoma detection in AS-OCT. Fig.~\ref{fig2} illustrates the pipeline of our proposed method. 


\subsection{Iris Segmentation Network with Wavelet Refinement Block}
\begin{figure}[t]
\includegraphics[width=\textwidth]{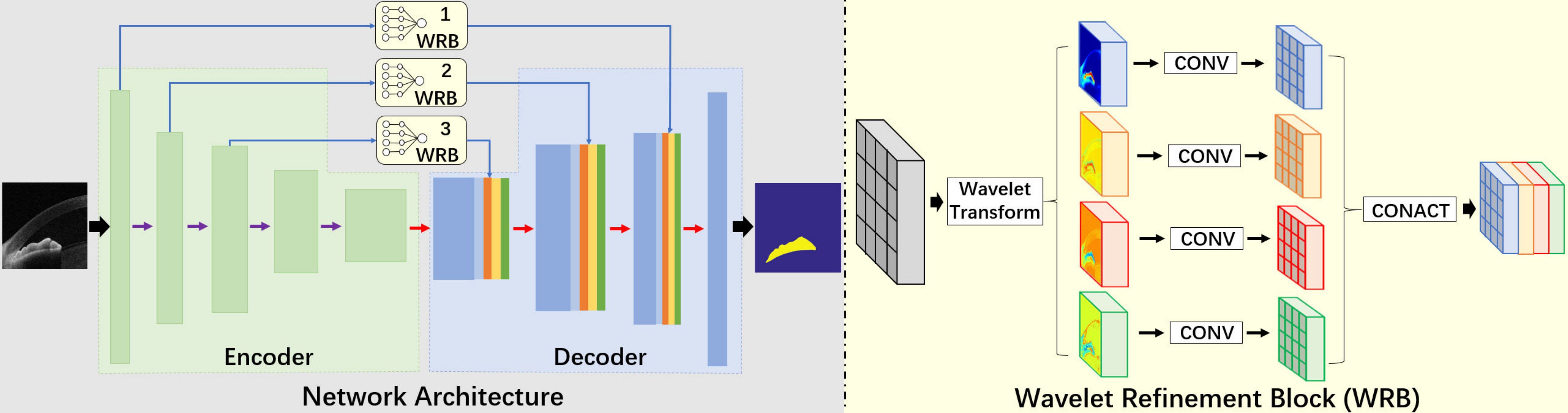}
\caption{Architecture of our segmentation model.} \label{fig3}
\vspace{-12pt}
\end{figure}

In~\cite{gu2019net,ronneberger2015u}, high resolution features from an encoder are combined  with decoder features using skip connection, which takes detailed information directly to the decoder to remedy  information loss due to pooling and convolutional operations. However, this operation also imports massive quantities of irrelevant information into the decoder, which disturbs and weakens the learning ability of networks. To address this issue, we introduce a new network component into the segmentation network, which we call a  wavelet refinement block (WRB). This is able to reduce the amount of  redundant information, while preserving  local details for the decoder. Fig.~\ref{fig3} illustrates the architecture of our segmentation network.

\vspace{-8pt}
\subsubsection{Discrete Wavelet Transform:} 
Given the input feature $X$, a 2D Discrete Wavelet Transform (DWT) with four convolutional filters - low-pass filter $f_{LL}$, and high-pass filters $f_{LH}$, $f_{HL}$, and $f_{HH}$ are performed  to decompose $X$ into four subband features, $Y_{LL}$, $Y_{LH}$, $Y_{HL}$, and $Y_{HH}$. Taking the Haar wavelet as an example, the four filters are defined as:
\begin{equation}\small
f_{LL}=\begin{bmatrix}1 & 1 \\1 & 1 \end{bmatrix},f_{LH}=\begin{bmatrix}-1 & -1 \\1 & 1 \end{bmatrix},f_{HL}=\begin{bmatrix}-1 & 1 \\-1 & 1 \end{bmatrix},f_{LL}=\begin{bmatrix}1 & -1 \\-1 & 1 \end{bmatrix}.
\label{eq_dwt}
\end{equation}
Note that all the convolutions above are performed with stride 2, yielding a subsampling of factor 2 along each spatial dimension. The DWT operation is defined as $Y_{LL}=(f_{LL}\otimes X)\downarrow_{2}$, $Y_{LH}=(f_{LH}\otimes X)\downarrow_{2}$, $Y_{HL}=(f_{HL}\otimes X)\downarrow_{2}$, and $Y_{HH}=(f_{HH}\otimes X)\downarrow_{2}$, where $\otimes$ denotes a convolution operator, and $\downarrow_{2}$ means the standard down-sampling operator with factor 2. 


\vspace{-8pt}
\subsubsection{Network Architecture:}
We utilize U-Net~\cite{ronneberger2015u} as the backbone, which consists of four encoder blocks which correspond to the first four blocks,  as shown in Fig.~\ref{fig3}. In order to restore boundary details, we insert three WRB, after the first three blocks to take local detail information to the decoder. Specifically, we use a Haar wavelet in Eq.~(\ref{eq_dwt}) to decompose the corresponding feature maps into four frequency subbands channel-wise, where each band is half-resolution of the input. It is worth noting that the low freq
uency band $Y_{LL}$ stores local averages of the input data: correspondingly, the high frequency bands, namely $Y_{LH}$, $Y_{HL}$, and $Y_{HH}$, encode details that are significant in recovering boundaries. We then  employ $1\times1$ convolution for each subband separately and cascade them with decoder feature maps. Allowing comparison with the skip connection that directly brings features from the encoder into the decoder in the general U-Net~\cite{ronneberger2015u}, our WRB reduces the introduction of redundant information while preserving  details, which makes our network more accurate and robust in predicting details.


\subsection{3D iris surface reconstruction and quantification}
At present, the gold standard for diagnostic angle assessment is observation of ACA by gonioscopy. In simpler terms, ophthalmologists move the gonioscope counterclockwise, making an annotation every $15^{\circ}$. In a similar manner, the AS-OCT automated scan obtains multiple consecutive radiant slices within a $15^{\circ}$ area, which can then be used to reconstruct a mesh of the iris surface in 3D. 


\vspace{-8pt}
\subsubsection{Surface Reconstruction:} 
\begin{figure}[t]
\includegraphics[width=\textwidth]{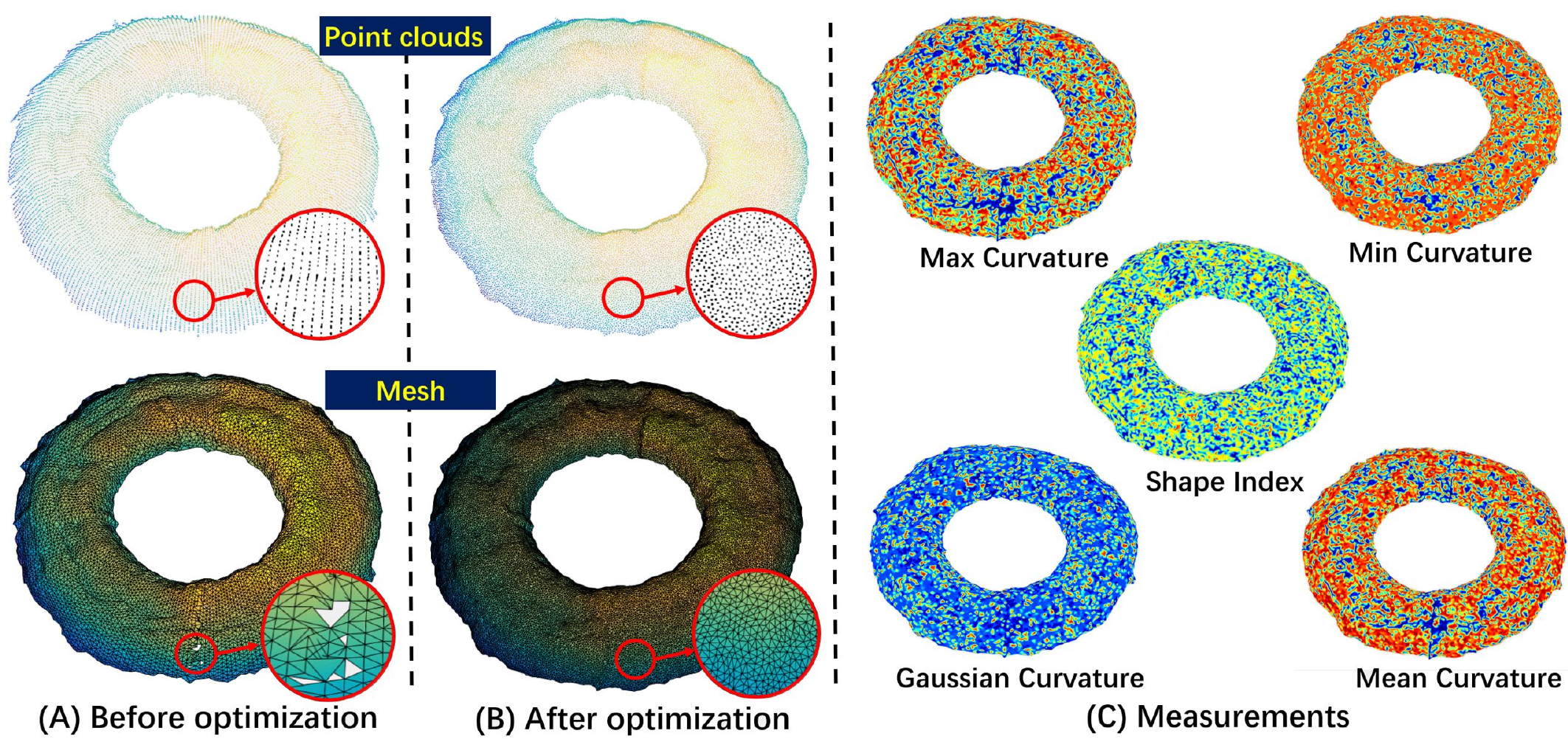}
\centering
\caption{Illustrative of iris 3D reconstruction and feature measurements. (A) Original point cloud and mesh generated by proposed 2D iris segmentation method. (B) Optimized point cloud and mesh. (C) Visualization of different measurements. } \label{fig4}
\end{figure}

Using  the previously obtained segmentation results, the upper boundaries of the iris are used to produce a 3D point cloud of the iris surface. These point clouds are nonuniform and sparse: nevertheless, the mesh generated is coarse and deficient of lacking local details. As demonstrated by the representative patch shown in Fig.~\ref{fig4} (A), distortions of the mesh lead to mispresentation of the iris surface. In addition, the geometrical changes in some regions are more dramatic than in others due to the existence of iris frill. This leads to a higher point density than in smooth areas, which leads to a low quality mesh, as shown in  Fig.~\ref{fig2} (C). 

To this end, we adapted a constrained Poisson-disk sampling~\cite{corsini2012efficient} of the coarse mesh  to refine the surface. This method produces a more uniform and dense point cloud,  while guaranteeing that objects of a certain size will be distributed according to the sampling scheme, without overlapping. In practice, an adaptive radius $r$ was utilized, to obtain a more precise representation of the point cloud while being as uniform as possible. Specifically, if the maximum curvature of a given point was larger than the global average, $r$ was set to $r_1$: otherwise, it was set to $r_2$. In our work, following empirical testing we set $r_1$ and $r_2$ to 6 and 10, respectively. 
Fig.~\ref{fig2} (D) and Fig.~\ref{fig4} (B) demonstrate the optimized point cloud and mesh, which are more effective in revealing geometrical details. 

\vspace{-8pt}
\subsubsection{Feature Extraction:}
As suggested by study~\cite{huang2015comparison}, the quantitative iris parameters, such as iris curvature, were independently associated to the degree of angle-closure progression. In consequence, 
after the reconstruction of the iris surface, we calculated the following curvature-related measures for the later diagnosis of angle-closure related diseases: principal curvatures, Gaussian curvature, mean curvature and shape index~\cite{Zhao2016}. 

It is worth noting that the $\textit {shape}$ $\textit {index}$ is introduced in order to capture the intuitive notion of `shape' locally and globally. The shape index $\mathscr{E}$ of each point may be defined as
\begin{equation} \small
{\mathscr{E}=\frac{2}{\pi}\arctan\frac{\mathscr{C}_2+\mathscr{C}_1}{\mathscr{C}_2-\mathscr{C}_1}},
\end{equation}
where ${\mathscr{C}_1}$ and ${\mathscr{C}_2}$ are the maximum and minimum curvatures of a point, respectively, and $\mathscr{E}\in[-1,1]$.
Unlike the curvature, the shape index is invariant to scaling of the shape, and it could give a simple measurement of the local shape - it can present the flat concave and convex regions significantly~\cite{Zhao2016}.



\begin{figure}[t]
\centering
\includegraphics[width=1\textwidth]{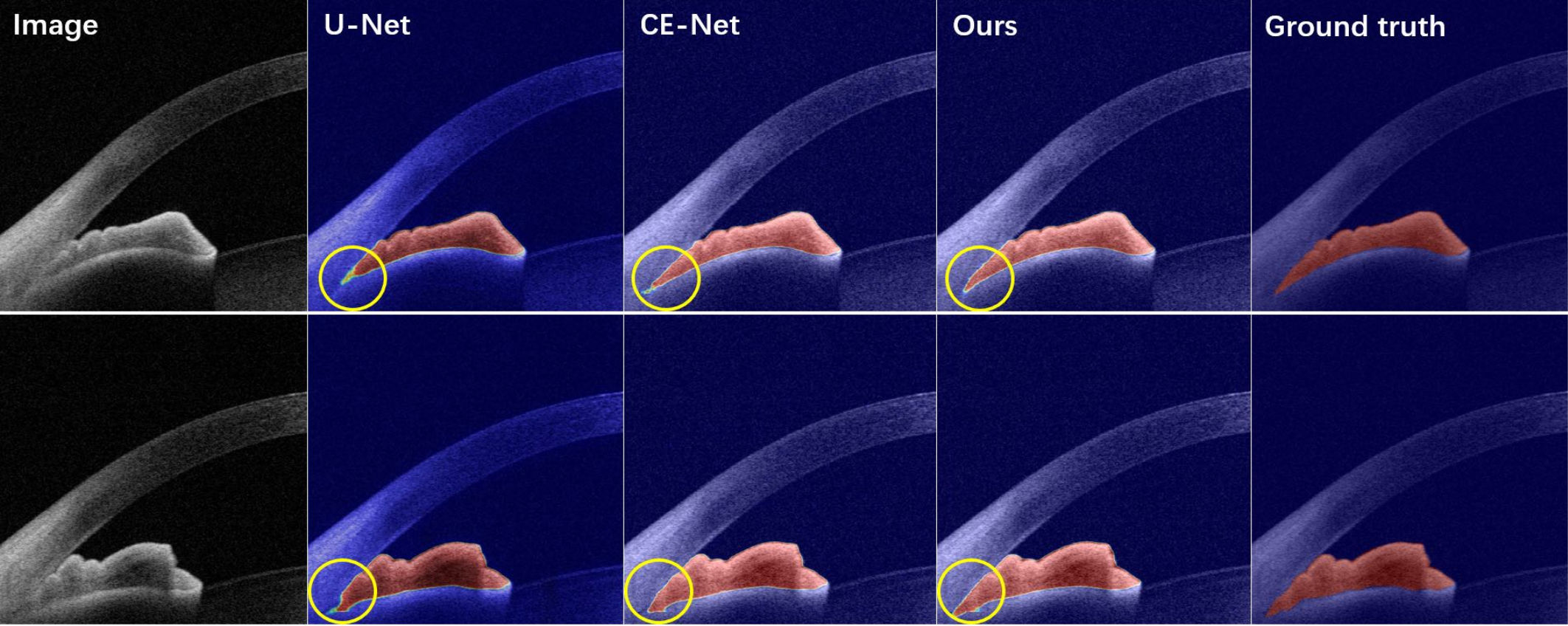}
\caption{Visualization of 2D iris segmentation results. From left to right: original image, and as obtained by U-Net~\cite{ronneberger2015u}, CE-Net~\cite{gu2019net}, our method, and ground truth.} \label{fig5}
\end{figure}

\section{Experimental Results}

In order to validate the effectiveness and superiority of the proposed method, we evaluated its individual components separately: first 2D iris segmentation, and then angle-closure glaucoma detection in 3D iris quantification.

\subsection{Evaluation of Iris Segmentation}

\vspace{-8pt}
\subsubsection{Data and metrics:}
A total of 100 AS-OCT images captured by CASIA-2 (Tomey Inc., Japan) from different subjects were collected. The iris regions were annotated by an image analysis expert and an ophthalmologist, and a consensus of their results was used as the final reference standard.  The dataset was divided equally into training and testing sets.  For a quantitative evaluation, we employed the following  metrics: accuracy (Acc), sensitivity (Sen) and dice coefficients (Dice) as region-level evaluation criteria; and root normalized mean square error (RNMSE) and Hausdorff distance (HD) as edge-level evaluation criteria. In addition, we also provided the trabecular iris contact (TIC) error score~\cite{fu2017segmentation}.


The segmentation performance of the proposed model was compared with the following  state-of-the-art segmentation methods: FCN~\cite{long2015fully}, U-Net~\cite{ronneberger2015u}, Segnet~\cite{badrinarayanan2017segnet}, and CE-Net~\cite{gu2019net}). As shown in Table~\ref{tab1}, our model achieved the best performance in terms of all metrics, with a single exception: its HD score is 0.01 lower than that of Segnet. 
Fig.~\ref{fig5} illustrates the iris segmentation results of the different methods over two sample images. Overall, the proposed method demonstrated that it could  correctly segment the iris region, when compared  with the corresponding manual annotations. From careful observation of the TIC regions, it was clear from visual inspection
that our method was better able to identify the iris root than the competing methods. This is because our method  introduces the wavelet refinement block to reduce redundancy in the image, so that our network may then  pay greater  attention to the salient context than other segmentation models. 

\begin{table}[t]\small
\centering
\caption{Performance in 2D iris segmentation of different methods.}\label{tab1}
\begin{tabular}{c|c|c|c|c|c|c}
	\hline\hline
	\multicolumn{1}{c|}{\multirow{2}{*}{\textbf{Method}}} &                                          \multicolumn{3}{c|}{\textbf{Region}}                                          &                   \multicolumn{2}{c|}{\textbf{Edge}}                   &       \multicolumn{1}{c}{\textbf{TIC}}       \\ \cline{2-7}
	                \multicolumn{1}{c|}{}                 & \multicolumn{1}{c|}{\textbf{Sen(\%)}} & \multicolumn{1}{c|}{\textbf{Dice(\%)}} & \multicolumn{1}{c|}{\textbf{Acc(\%)}} & \multicolumn{1}{c|}{\textbf{RNMSE}} & \multicolumn{1}{c|}{\textbf{HD}} & \multicolumn{1}{c}{\textbf{TIC Err.(Pixel)}} \\ \hline
	              FCN~\cite{long2015fully}                &                 84.63                 &                 87.01                  &                 97.24                 &               0.0601                &               6.60               &                    21.77                     \\
	            U-Net~\cite{ronneberger2015u}             &                 88.19                 &                 91.84                  &                 99.26                 &               0.0543                &               4.64               &                    19.63                     \\
	       Segnet~\cite{badrinarayanan2017segnet}         &                 91.09                 &                 92.70                  &                 99.32                 &               0.0619                &          \textbf{4.31}           &                    17.65                     \\
	               CE-Net~\cite{gu2019net}                &                 95.74                 &                 94.89                  &                 99.45                 &               0.0512                &               4.41               &                    14.43                     \\ \hline
	                      Proposed                        &            \textbf{96.46}             &             \textbf{95.21}             &            \textbf{99.48}             &           \textbf{0.0504}           &               4.32               &                \textbf{13.24}                \\ \hline\hline
\end{tabular}
\end{table}

\subsection{Evaluation of Angle-Closure Glaucoma Classification}

In this experiment, we first reconstructed 3D iris surfaces using different segmentation methods:  FCN~\cite{long2015fully}, U-Net~\cite{ronneberger2015u}, Segnet~\cite{badrinarayanan2017segnet} and CE-Net~\cite{gu2019net}. We then used measurements obtained in Section 2.2 as the input of a 3D classification model, PointNet~\cite{qi2017pointnet}, to classify the glaucoma subjects into cases of open angle and angle-closure glaucoma types, respectively.

\vspace{-8pt}
\subsubsection{Data and metrics:} A total of 42 AS-OCT volumes were captured from 42 eyes, with  each volume containing 128 AS-OCT images. A senior ophthalmologist made an annotation (determining open or angle-closure glaucoma) from gonioscopic examination of every $15^{\circ}$ segment ACA of each  eye, yielding 24 annotations for a single eye, resulting in a total of 1008
annotations for each dataset (504 open-angle and 504 angle-closure). In light of this, we partitioned the 42 automatically-generated 3D iris shapes into 1008 sub-surfaces by dividing each shape into 24 $15^{\circ}$ segments. We trained using 80\% of these sub-surfaces and reserved 20\% as a testing set. We employed the metrics of  \textit{Acc}, \textit{Sen}, \textit{Spe}, and area under ROC curve (AUC) to measure the final angle-closure glaucoma classification performance.

\vspace{-8pt}
\subsubsection{Results:} Table~\ref{tab2} reports the classification performances using features extracted from the 3D iris reconstruction by the different segmentation methods. It may be seen that clearly significant margins of improvement in classification results were achieved when the proposed method is compared with the other state-of-art segmentation models. For example, our method  exhibits a large advantage over  FCN by increases in  $Acc$ and $Sen$ of about  8.33\% and 10.5\%, respectively.
Comparatively, our method reduces noise and other redundancies in AS-OCT, thereby allow  the decoder to concentrate on high-level context, such as the TIC region, which is more beneficial for disease detection.

\vspace{-8pt}
\subsubsection{Effectiveness of optimization step:} In addition, we also demonstrate how the classification result benefits from the point cloud optimization process. We may observe that the generated 3D iris surface seen in Fig.~\ref{fig4} A, suffers from distortion without the subsequent optimization step optimization step (Fig.~\ref{fig4} B), and this scenario leads to incorrect curvature estimation, which will further compromise the accuracy of classification. This finding is evidenced in Table~\ref{tab2}: without the point cloud optimization step, the classification results show decrements of  3.1\%, 2.1\%, and 4.2\% for ACC, Sen and Spe, respectively.

\vspace{-8pt}
\subsubsection{Comparison of 2D and 3D features:}
All existing angle-closure glaucoma classification methods are accomplished using 2D features obtained  from a single AS-OCT slice.  To further verify whether  the features extracted from a 3D iris surface obtained by our method could improve classification performance, we compared the proposed method to the conventional approaches that use 2D feature representation: histograms of oriented gradients (HOG) features\cite{xu2012anterior} with liner Support Vector Machine (SVM), AlexNet~\cite{krizhevsky2012imagenet}, VGG~\cite{simonyan2014very}, and ResNet~\cite{he2016}. The classification performances of these methods are reported in Table~\ref{tab2}. It may be seen that by enabling the use of features extracted from a reconstructed 3D iris surface, the 3D classification network, PointNet, achieved the best performances in terms of Acc, Spe, and AUC. This confirms that making use of  3D features is more helpful in improving the accuracy of angle-closure glaucoma classification than using 2D features alone.

\begin{table}[!t]\small
    \centering
    \caption{Angle-closure glaucoma classification results obtained by different methods.}\label{tab2}
\setlength{\tabcolsep}{1.5mm}{ 
    \begin{tabular}{l|cccc}
    	\hline\hline
    	\textbf{}                         & \textbf{Acc(\%)} & \textbf{Sen(\%)} & \textbf{Spe(\%)} & \textbf{AUC(\%)} \\ \hline
    	2D: HOG+SVM                        &      85.01       &      74.33       &      95.90       &      91.92       \\
    	2D: ResNet34                       &      96.87       &      97.91       &      94.79       &      99.52       \\
    	2D: AlexNet                        &      91.99       &  \textbf{99.60}  &      84.37       &      93.81       \\
    	2D: VGG16                          &      94.18       &      96.48       &      90.13       &      98.95       \\ \hline
    	3D: FCN  + PointNet                   &      90.10       &      88.54       &      91.66       &      97.55       \\
    	3D: U-Net + PointNet                  &      92.70       &      89.58       &      95.83       &      98.82       \\
    	3D: Segnet + PointNet                 &      94.79       &      96.87       &      92.70       &      99.01       \\ \hline
    	Our WRB + PointNet                &      95.31       &      96.87       &      93.75       &      99.17       \\
    	Our WRB + Optimization + PointNet &  \textbf{98.43}  &      98.95       &  \textbf{97.91}  &  \textbf{99.83}  \\ \hline\hline
    \end{tabular}
}
\end{table}

\section{Conclusion}

Existing methods to identify gonioscopic angle-closure have focused solely on the extraction of features from 2D slices, which are less satisfactory  for the identification  of angle-closure glaucoma subtypes. 
In this work, we have developed  a novel framework for reconstruction and quantification of 3D iris surface from AS-OCT imagery.
This is for the first time that a comprehensive surface-based framework has been applied to model and analyze 3D iris from AS-OCT.  
The high evaluation performance in segmentation experiments shows the ability and robustness of our models in extracting iris features from single AS-OCT slices. Feature analysis and glaucoma screening have then  been performed based on 3D iris reconstruction, which show the high effectiveness of our approach. The proposed framework opens the possibility for further investigation of  AS-OCT from a new perspective. 

\bibliographystyle{splncs}
\bibliography{library}

\begin{thebibliography}{10}

\bibitem{Chansangpetch2018}
Chansangpetch, S., Rojanapongpun, P., Lin, S.C.:
\newblock Anterior segment imaging for angle closure.
\newblock American journal of ophthalmology \textbf{188} (2018)  xvi--xxix

\bibitem{Ang2018}
Ang, M., Baskaran, M.,  et~al.:
\newblock Anterior segment optical coherence tomography.
\newblock Progress in retinal and eye research \textbf{66} (2018)  132--156

\bibitem{Xu2019_AJO}
Xu, B.Y., Chiang, M.,  et~al.:
\newblock Deep learning classifiers for automated detection of gonioscopic
  angle closure based on anterior segment oct images.
\newblock American journal of ophthalmology \textbf{208} (2019)  273--280

\bibitem{fu2017segmentation}
Fu, H., Xu, Y.,  et~al.:
\newblock Segmentation and quantification for angle-closure glaucoma assessment
  in anterior segment oct.
\newblock IEEE transactions on medical imaging \textbf{36}(9) (2017)
  1930--1938

\bibitem{Fu2018_MICCAI}
Fu, H., Xu, Y.,  et~al.:
\newblock {Multi-Context Deep Network for Angle-Closure Glaucoma Screening in
  Anterior Segment OCT}.
\newblock In: International Conference on Medical image computing and
  computer-assisted intervention. (2018)  356--363

\bibitem{Fu2019_AJO}
Fu, H., Baskaran, M.,  et~al.:
\newblock A deep learning system for automated angle-closure detection in
  anterior segment optical coherence tomography images.
\newblock American journal of ophthalmology \textbf{203} (2019)  37--45

\bibitem{Fu2019TCb}
Fu, H., Xu, Y.,  et~al.:
\newblock {Angle-Closure Detection in Anterior Segment OCT Based on Multilevel
  Deep Network}.
\newblock IEEE Transactions on Cybernetics (2020)

\bibitem{wang2010quantitative}
Wang, B., Sakata, L.M.,  et~al.:
\newblock Quantitative iris parameters and association with narrow angles.
\newblock Ophthalmology \textbf{117}(1) (2010)  11--17

\bibitem{huang2015comparison}
Huang, J., Wang, Z., Wu, Z., Li, Z., Lai, K., Ge, J.:
\newblock Comparison of ocular biometry between eyes with chronic primary
  angle-closure glaucoma and their fellow eyes with primary angle-closure or
  primary angle-closure suspect.
\newblock Journal of glaucoma \textbf{24}(4) (2015)  323--327

\bibitem{ni2014anterior}
Ni~Ni, S., Tian, J., Marziliano, P., Wong, H.T.:
\newblock Anterior chamber angle shape analysis and classification of glaucoma
  in ss-oct images.
\newblock Journal of ophthalmology \textbf{2014} (2014)

\bibitem{shang2019tensor}
Shang, Q., Zhao, Y.,  et~al.:
\newblock Automated iris segmentation from anterior segment oct images with
  occludable angles via local phase tensor.
\newblock In: Annual International Conference of the IEEE Engineering in
  Medicine and Biology Society (EMBC), IEEE (2019)  4745--4749

\bibitem{cho2017evaluation}
Cho, H.k., Ahn, D., Kee, C.:
\newblock Evaluation of circumferential angle closure using iridotrabecular
  contact index after laser iridotomy by swept-source optical coherence
  tomography.
\newblock Acta ophthalmologica \textbf{95}(3) (2017)  e190--e196

\bibitem{ronneberger2015u}
Ronneberger, O., Fischer, P., Brox, T.:
\newblock U-net: Convolutional networks for biomedical image segmentation.
\newblock In: International Conference on Medical image computing and
  computer-assisted intervention, Springer (2015)  234--241

\bibitem{gu2019net}
Gu, Z., Cheng, J.,  et~al.:
\newblock Ce-net: context encoder network for 2d medical image segmentation.
\newblock IEEE transactions on medical imaging \textbf{38}(10) (2019)
  2281--2292

\bibitem{corsini2012efficient}
Corsini, M., Cignoni, P., Scopigno, R.:
\newblock Efficient and flexible sampling with blue noise properties of
  triangular meshes.
\newblock IEEE transactions on visualization and computer graphics
  \textbf{18}(6) (2012)  914--924

\bibitem{Zhao2016}
Zhao, Y., Liu, Y.,  et~al.:
\newblock Region-based saliency estimation for 3d shape analysis and
  understanding.
\newblock Neurocomputing \textbf{197} (2016)  1--13

\bibitem{long2015fully}
Long, J., Shelhamer, E., Darrell, T.:
\newblock Fully convolutional networks for semantic segmentation.
\newblock In: CVPR. (2015)  3431--3440

\bibitem{badrinarayanan2017segnet}
Badrinarayanan, V., Kendall, A., Cipolla, R.:
\newblock Segnet: A deep convolutional encoder-decoder architecture for image
  segmentation.
\newblock IEEE transactions on pattern analysis and machine intelligence
  \textbf{39}(12) (2017)  2481--2495

\bibitem{qi2017pointnet}
Qi, C.R., Su, H., Mo, K., Guibas, L.J.:
\newblock Pointnet: Deep learning on point sets for 3d classification and
  segmentation.
\newblock In: CVPR. (2017)  652--660

\bibitem{xu2012anterior}
Xu, Y., Liu, J.,  et~al.:
\newblock Anterior chamber angle classification using multiscale histograms of
  oriented gradients for glaucoma subtype identification.
\newblock In: Annual International Conference of the IEEE Engineering in
  Medicine and Biology Society, IEEE (2012)  3167--3170

\bibitem{krizhevsky2012imagenet}
Krizhevsky, A., Sutskever, I., Hinton, G.E.:
\newblock Imagenet classification with deep convolutional neural networks.
\newblock In: Advances in neural information processing systems. (2012)
  1097--1105

\bibitem{simonyan2014very}
Simonyan, K., Zisserman, A.:
\newblock Very deep convolutional networks for large-scale image recognition.
\newblock arXiv preprint arXiv:1409.1556 (2014)

\bibitem{he2016}
He, K., Zhang, X., Ren, S., Sun, J.:
\newblock Deep residual learning for image recognition.
\newblock In: CVPR. (2016)  770--778

\end{thebibliography}

\end{document}